\documentclass{Interspeech}

\interspeechcameraready

\title{Incorporating Linguistic Constraints from External Knowledge Source for Audio-Visual Target Speech Extraction}

\author[affiliation={1}]{Wenxuan}{Wu}
\author[affiliation={2,3}]{Shuai}{Wang}
\author[affiliation={1}]{Xixin}{Wu}
\author[affiliation={1}]{Helen}{Meng}
\author[affiliation={2,4}]{Haizhou}{Li}

\affiliation{Department of SEEM}{The Chinese University of Hong Kong}{Hong Kong SAR, China}
\affiliation{SRIBD, School of Data Science}{The Chinese University of Hong Kong, Shenzhen}{China}
\affiliation{School of Intelligence Science and Technology}{Nanjing University, Suzhou}{China}
\affiliation{Department of ECE}{National University of Singapore}{Singapore}

\email{\{wwu, wuxx, hmmeng\}@se.cuhk.edu.hk, shuaiwang@nju.edu.cn, haizhouli@cuhk.edu.cn }


\usepackage[symbol]{footmisc}

\usepackage{algorithm}
\usepackage[noend]{algpseudocode}
 \usepackage{multirow}
 \usepackage[normalem]{ulem} 
 \usepackage{hyperref} 
 \usepackage{pifont} 
\usepackage{adjustbox}
\usepackage{graphicx}
\usepackage{footnote}

\begin{document}

\maketitle

\begin{abstract}
\let\thefootnote\relax\footnotetext{Corresponding author: Shuai Wang}
Audio-visual target speaker extraction (AV-TSE) models primarily rely on target visual cues to isolate the target speaker's voice from others. We know that humans leverage linguistic knowledge, such as syntax and semantics, to support speech perception.  Inspired by this, we explore the potential of pre-trained speech-language models (PSLMs) and pre-trained language models (PLMs) as auxiliary knowledge sources for AV-TSE. In this study, we propose incorporating the linguistic constraints from PSLMs or PLMs for the AV-TSE model as additional supervision signals. Without introducing any extra computational cost during inference, the proposed approach consistently improves speech quality and intelligibility. Furthermore, we evaluate our method in multi-language settings and visual cue-impaired scenarios and show robust performance gains.

\end{abstract}

\section{Introduction}
Audio-visual target speaker extraction (AV-TSE) simulates the human ability to extract the target speaker's utterance in a cocktail party scenario using visual cues. Most existing studies focus on improving the audio-visual fusion mechanisms \cite {muse, av-sepformer,wu2024target_cvpr,ijcnn,wu2025c} or addressing visual cue-impaired scenarios \cite {ImagineNET,Wang2024RestoringSL}. Despite the progress, the role of external knowledge has often been overlooked. A previous study suggests that when humans listen to words with similar semantics, the same neural units in the human auditory area will be activated \cite{nature}. In other words, linguistic knowledge may also serve as a useful cue for human's auditory perception. The linguistic knowledge functions as a prior and could be pre-trained or pre-loaded from an external knowledge source. 

Such biological insight motivates us to incorporate external linguistic knowledge into AV-TSE systems. Pretrained language models (PLMs) and pretrained speech-language models (PSLMs) encode rich knowledge from massive unlabeled text and speech data through self-supervised learning, making them promising knowledge sources \cite{HuBERT, mengicassp2025}. While previous studies attempted PSLMs to improve TSE performance, most methods directly inject the PSLM features into TSE systems~\cite{ Peng2024TargetSE,Probing_ssl_tse}. However, this approach poses several risks: First, PSLMs capture both task-relevant and task-irrelevant knowledge~\cite{yang2024large}, feeding irrelevant knowledge into the extractor may degrade performance or introduce unwanted interference \cite{kt_icml}. 
Second, features derived from PSLMs might not perfectly align with intermediate features in TSE models, requiring additional fusion or alignment mechanisms, which may introduce additional parameter overhead \cite{Peng2024TargetSE,wu2024target_cvpr,ijcnn}.
Third, the expensive computation cost of PSLM/PLMs makes it infeasible for edge devices when such pretrained models are involved in the inference process.
  
To address these challenges, we propose to extract linguistic knowledge from PSLMs or PLMs as additional supervision signals during the optimization process.
Similar approaches have been explored in some speech-related tasks.
In~\cite{han23_interspeech}, linguistic knowledge from PLMs is distilled into ASR models to enhance speech recognition performance. Additionally, recent studies on neural speech codecs~\cite{zhang2023speechtokenizer,Guo2024LSCodecLA, ALMTOKENIZER} have leveraged PSLMs to improve the modeling of linguistic information, demonstrating the effectiveness of integrating pretrained knowledge into speech compression frameworks. Despite its success in other speech applications, for AV-TSE it remains underexplored. In this study, we investigate different PSLMs and PLMs as external knowledge sources and evaluate their effectiveness in enhancing AV-TSE systems. Key contributions are as follows,

\begin{itemize}
\item We systematically investigate the effectiveness of different pretrained language models (PLMs) and pretrained speech-language models (PSLMs) as knowledge sources for AV-TSE and design corresponding adapter modules to effectively integrate their linguistic knowledge.
\item We comprehensively evaluate our method's robustness across multiple languages and challenging conditions, such as visual cue-impaired scenarios.
\item Our method consistently enhances speech extraction quality in terms of signal similarity, speech intelligibility, and semantic coherence, demonstrating the effectiveness of using linguistic constraints for AV-TSE.
\end{itemize}

\section{Methods}
\subsection{Problem Formulation}
Given a mixture speech signal $x$ and the target speaker's visual cue $v$, the goal of an AV-TSE system $f_\theta$ is to extract a clean speech signal $\hat{y}$ that approximates the ground truth target speech $y$ via $\hat{y}=f_\theta(x, v)$. In most conventional AV-TSE systems, the model parameter $\theta$ can be obtained by optimizing the SI-SDR loss $\mathcal{L}_\text{SI-SDR}$ as
\vspace{-3pt}
\begin{equation}
    \theta^* = \arg \min_{\theta} \mathcal{L}_{\text{SI-SDR}}(f_{\theta}(x, v), y)
\end{equation}

Given an external knowledge source (such as a pretrained language model (PLM) or a pretrained speech-language model (PSLM)), we aim to introduce an additional supervision term to incorporate linguistic constraint into the optimization process. To achieve this, we modify the objective function as follows:

\begin{equation}
    \theta^{**} = \arg \min_{\theta} \underbrace{\alpha\mathcal{L}_{\text{SI-SDR}}(f_{\theta}(x, v), y)}_{\text{Signal Reconstruction}} +  \underbrace{\beta\mathcal{L}_\text{  LC}(\hat{z}, z)}_{\text{Linguistic Constraint }}
\end{equation}
where $\hat{z}$ is the linguistic representation obtained from the extracted speech $\hat{y}$, and $z$ is the target representation obtained from a pretrained knowledge model.

Note that when a PSLM is used as an external knowledge source, $z$ represents a sequence of latent vectors derived from the ground truth speech $y$, and $\hat{z}$ is a sequence of the same shape but derived from the predicted speech $\hat{y}$. Conversely, when a PLM serves as an external knowledge source, $z$ is a compressed vector that captures the global linguistic information of the transcript of $y$, while $\hat{z}$ is a vector of the same dimensionality as $z$, but obtained by averaging the latent sequence extracted from the PSLM. In this case, as illustrated in Figure.\ref{fig:kt}, additional adapters are required, assuring that $z$ and $\hat{z}$ are in the same latent space.


\subsection{Incorporation of Linguistic Constraints}


The linguistic knowledge  is encoded either in a PSLM or a PLM:
\begin{itemize}
    \item A PSLM accepts audio as input and provides phoneme-level semantic and acoustic contextual features.
    \item A PLM accepts text as input and provides contextualized linguistic information at the word and sentence level.
\end{itemize}

\begin{figure}[!htb]
\centering
\includegraphics[scale=0.35]{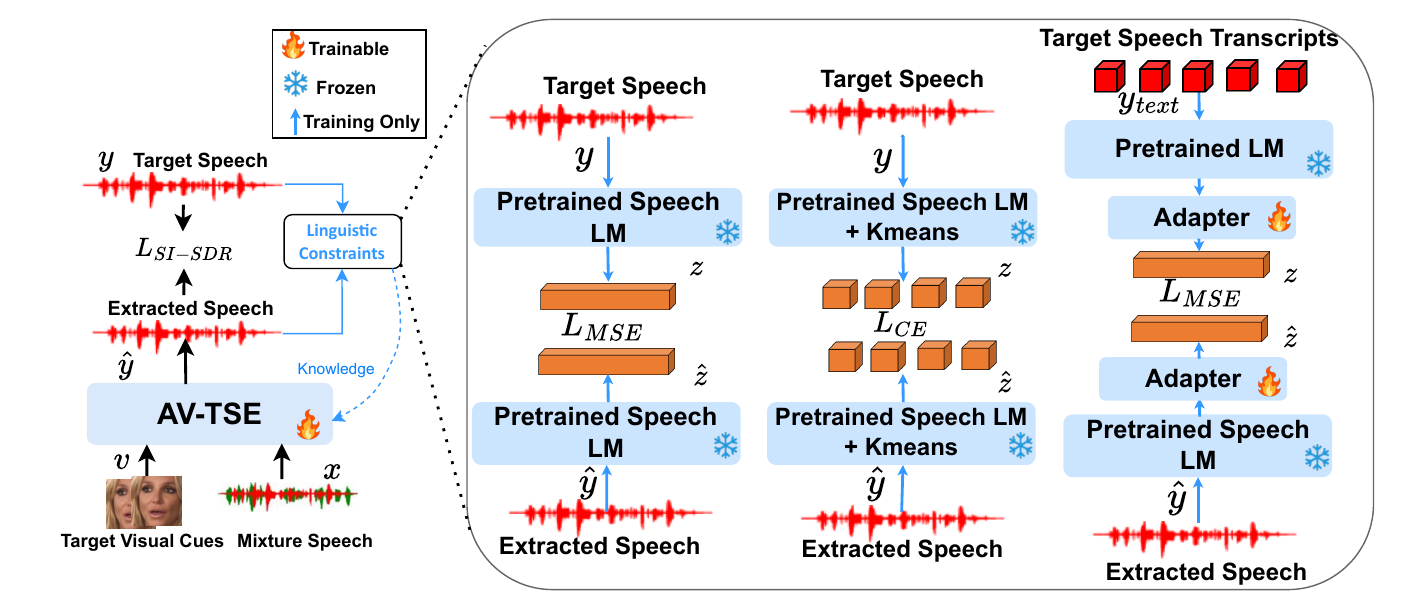}
\caption{AV-TSE model with linguistic constraints. From left to right: continuous features and discrete tokens of PSLM,  continuous features of PLM.}
\label{fig:kt}
\end{figure}

A high-level illustration of the AV-TSE system incorporated with linguistic constraints is shown in Figure~\ref{fig:kt}, where different forms of integration are also depicted.

We explore two ways to incorporate PSLMs: continuous features and discrete tokens after quantization. For continuous features, we utilize feature representations extracted from the 24th layer of HuBERT-Large (HuBERT-L-L24) ~\cite{HuBERT}, the 24th layer of WavLM-Large (WavLM-L-L24) ~\cite{WavLM}, and the 12th layer of WavLM-Base (WavLM-B-L12). We employ the Mean Squared Error (MSE) loss to minimize discrepancies between the extracted features and the corresponding knowledge representations.  For discrete tokens, we utilize tokenized representations extracted from HuBERT-L-L24. The Cross-Entropy (CE) loss is adopted to minimize the discrepancy between the predicted and reference token distributions.

To incorporate linguistic constraints from PLMs, we employ the 12th layer of RoBERTa-Base (RoBERTa-B-L12)~\cite{Liu2019RoBERTaAR} and utilize the target speech transcript as input to obtain the corresponding linguistic feature representation as illustrated in Figure.~\ref{fig:kt}. For extracted speech $\hat{y}$, we could not use the same PLMs due to modality mismatch, instead, we adopt a PSLM to encode $\hat{y}$ and introduce two adapters to bridge the latent space shift, then we calculate the MSE loss between two projected embeddings to enhance linguistic constraints.

\begin{table*}[ht]
\centering
\caption{The study of incorporating linguistic constraints from different knowledge sources, that is evaluated on the core test set. ``EP'' denotes whether extra parameters are required during inference.}
\resizebox{0.9\textwidth}{!}{%
\begin{tabular}{c|c|c c c c c c}  
\hline
\multirow{2}{*}{\textbf{Knowledge Source}} & \multirow{2}{*}{\textbf{EP}} & \multicolumn{6}{c}{\textbf{Core Test Set}} \\
\cline{3-8}
 & & \textbf{SI-SDR $(\uparrow)$} & \textbf{SI-SDRi $(\uparrow)$} & \textbf{SDR $(\uparrow)$} & \textbf{PESQ $(\uparrow)$} & \textbf{STOI $(\uparrow)$} & \textbf{SpeechBERTScore $(\uparrow)$} \\
\hline
- (baseline)  & \ding{55} & 12.03 & 12.07 & 13.35 & 2.78 & 0.88 & 0.8247 \\ 
 MAR* \cite{wu2024target_cvpr}  &  \ding{51}  & 13.18 & 13.23 & 13.46 & 3.05 & 0.89 & 0.8563 \\  
AVSepChain \cite{avsepchain} &  \ding{51}  & - & 13.60 & - & 2.72 & - & - \\
CTC ** &  \ding{55}  & 13.41 & 13.46 & 13.73 & 3.00 & 0.90 & 0.8521 \\ 
\hline
HuBERT-L-L24 (token) &  \ding{55}  & 13.01 & 13.06 & 13.33 & 2.94 & 0.89 & 0.8442 \\ 
HuBERT-L-L24 &  \ding{55}  & 13.05 & 13.10 & 13.38 & 2.94 & 0.89 & 0.8438 \\ 
WavLM-B-L12 &  \ding{55}  & 13.02 & 13.07 & 13.35 & 2.94 & 0.89 & 0.8439 \\ 
WavLM-L-L24 &  \ding{55}  & 13.34 & 13.38 & 13.67 & 3.00 & 0.90 & 0.8510 \\ 
\hline
RoBERTa-B-L12 &  \ding{55}  & \textbf{13.60} & \textbf{13.65} & \textbf{13.93} & \textbf{3.05} & \textbf{0.90} & \textbf{0.8567} \\ 
\hline
\end{tabular}}

\vspace{0.5em} 
\footnotesize 
* MAR and ** CTC don't rely on external pretrained models
\label{multi_test}
\end{table*}



\section{Experimental Setups}
\subsection{Dataset}
In this study, several experimental settings are considered:
\begin{itemize}
\item \textbf{Training Set:}
A two-speaker mixture training set is simulated following previous work~\cite{muse, ImagineNET, av-sepformer, wu2024target_cvpr}. The training set contains  $800$ speakers with $20,000$,   utterances. Speakers appearing in the test sets are excluded from the training data and SNR of interfering utterances are set randomly sampled from -10db to 10db.  Among training utterances, around 76\% English (EN), 6.4\%  French (FR), 0.5\% Italian (IT), 1.4\% Spanish
(ES), 0.14\% Portuguese (PT). Also includes others like 1.2\% Dutch (NL) 7.7\% German (DE), etc.  All speech segments are clipped to 4 seconds.

\item \textbf{Test Set:} Multiple test sets are designed to evaluate the proposed methods under diverse conditions:
\begin{itemize}
    \item \textbf{Core:}  
    A two-speaker mixture set is created following~\cite{muse, ImagineNET, av-sepformer, wu2024target_cvpr}, with 118 randomly selected speakers and 3,000 simulated mixtures. Each mixture contains speakers of the same or different languages, with EN comprising the largest proportion. During inference, all segments are set to 6 seconds, shorter utterances are zero-padded.
    
    \item \textbf{Monolingual:}  
    Five two-speaker mixture test sets are generated for IT, PT, EN, ES, and FR. Each mixture contains speakers from the same language. Due to insufficient test data for non-majority languages in VoxCeleb2~\cite{voxceleb2}, 118 speakers are selected from the development set to simulate 2,000 monolingual mixtures per language, all segments are set to 6 seconds, shorter utterances are zero-padded.
    \item \textbf{Visual Cue Impaired:}  
Three scenarios are simulated:   partial occlusion, low resolution, and full visual missing, the first two following~\cite{Hong2023WatchOL}, where we randomly select an obstacle for partial occlusion and apply a Gaussian blur or Gaussian noise to represent low resolution. Full visual missing frames are set to zero value. For each scenario, the impairment ratio ranges from 0\% to 100\%,  uniformly distributed.

    \item \textbf{Cross-Domain:}  
    Generalization is assessed by evaluating the model on the Lip Reading Sentences 3 (LRS3) test set~\cite{Afouras2018LRS3TEDAL}, following the test simulation protocol in~\cite{muse}.
    
\end{itemize}
\end{itemize}
\subsection{Evaluation Metric}
Speech quality is evaluated using the Scale-Invariant Signal-to-Distortion Ratio (SI-SDR)~\cite{SDR} and its improvement (SI-SDRi). Perceptual quality and intelligibility are assessed with the Perceptual Evaluation of Speech Quality (PESQ)~\cite{pesq} and the Short-Term Objective Intelligibility (STOI)~\cite{stoi}. To measure semantic similarity between extracted speech and ground truth, SpeechBERTScore with HuBERT Base~\cite{saeki2024speechbertscore} is employed. Higher values in all metrics indicate better performance.
\subsection{Implementation Details}
AV-Sepformer~\cite{av-sepformer} is selected as the AV-TSE backbone. For adapter modules utilized for PLMs, we employed two fully connected layers. We follow \cite{mousavi24_interspeech} to discrete HuBERT continuous features with 500 cluster centers.
A two-stage training framework is employed.  
In the first stage, AV-Sepformer is pretrained for 100 epochs. Training is halted if no validation improvement is observed for 6 epochs, with the learning rate halved after 3 consecutive epochs without improvement. Only $L_\text{SI-SDR}$ is used as the training objective. 
In the second stage, training continues with the same settings, incorporating both $L_\text{SI-SDR}$ and $L_\text{LC}$. The weighting factors $\alpha$ and $\beta$ are set to 1 and 10, respectively. 
 \vspace{-5pt} 
\section{Results and Analysis}

In this section, linguistic knowledge from different pretrained models is evaluated and analyzed on various test sets. For a more comprehensive comparison, we also include two systems that do not explicitly incorporate pretrained models as external knowledge source but have been validated in previous studies as beneficial for enhancing the model’s linguistic modeling capability.
The first is the Mask-And-Recover (MAR)  \cite{wu2024target_cvpr} strategy, which incorporates inter- and intra-modality contextual correlations into AV-TSE models by randomly masking mixture and recovering it with MAR block. The original paper implemented the MAR strategy on MuSE \cite{muse}, but it could also be adapted to other backbones. In this study, we implemented the MAR strategy on AV-Sepformer and retrained it using the same training settings. Another is jointly optimizing AV-Sepformer with Connectionist Temporal Classification (CTC) loss \cite{GravesConnectionistTC}, which assists in the temporal alignment between acoustic features and textual logits. In this study, we use Wav2Vec2ForCTC \cite{wav2vec2} to implement CTC between target speech transcripts and AV-Sepformer outputs. 

 Additionally, a system explicitly integrates PSLM into AVSepformer to enhance model’s linguistic modeling capability. AVSepChain explicitly integrates HuBERT and AV-HuBERT with AVSepformer, focusing on visual-speech semantic matching \cite{avsepchain}. We also use AVSepChain results from the original paper for comparison.


\subsection{Main experiments on the Core Test Set}
Results on the core test set are shown in Table \ref{multi_test}. In general, optimizing with language constraints from any of the pre-trained knowledge sources consistently improves performance compared to the baseline. Here are some key observations:
 
\begin{table*}[ht]
\centering
\caption{The study of incorporating linguistic constraints from different knowledge sources, that is evaluated on the test sets of different languages. The percentage in parentheses indicates the proportion of data for each language in the training set, as mentioned in Section 3.1. }
\vspace{-5pt}
  \resizebox{0.9\textwidth}{!}{
\begin{tabular}
{c|r r|r r|r r|r r|r r}   
\hline
\multirow{2}{*}{\textbf{Knowledge Source}} & \multicolumn{2}{c|}{\textbf{EN} (76\%)} & \multicolumn{2}{c|}{\textbf{PT}(0.14\%)} & \multicolumn{2}{c|}{\textbf{ES} (1.4\%)} & \multicolumn{2}{c|}{\textbf{IT} (0.5\%)} & \multicolumn{2}{c}{\textbf{FR} (6.4\%)} \\ 
\cline{2-11}
 & \textbf{SI-SDRi } & \textbf{PESQ  } & \textbf{SI-SDRi  } & \textbf{PESQ  } & \textbf{SI-SDRi  } & \textbf{PESQ } & \textbf{SI-SDRi  } & \textbf{PESQ  } & \textbf{SI-SDRi  } & \textbf{PESQ  }\\ 
\hline
- (baseline)          & 11.79 & 2.72 & 11.34 & 2.71 & 11.38 & 2.74 & 11.67 & 2.76 & 11.80 & 2.56 \\ 
 MAR* \cite{wu2024target_cvpr}              & 12.72 & 2.90 & 12.25 & 2.89 & 12.20 & 2.90 & 12.51 & 2.93 & 12.93 & 2.72 \\ 
 CTC**           & 13.03 & 2.92 & 12.60 & 2.90 & 12.59 & 2.94 & 12.84 & 2.94 & 13.18 & 2.77 \\
 \hline
HuBERT-L-L24 (token) & 12.63 & 2.86 & 12.26 & 2.85 & 12.24 & 2.89 & 12.47 & 2.89 & 12.84 & 2.71 \\ 
 HuBERT-L-L24     & 12.65 & 2.86 & 12.20 & 2.84 & 12.30 & 2.88 & 12.49 & 2.89 & 12.87 & 2.71 \\ 
 WavLM-B-L12 & 12.66 & 2.86 & 12.09 & 2.83 & 12.22 & 2.88 & 12.37 & 2.87 & 12.77 & 2.70 \\ 
 WavLM-L-L24      & 12.99& 2.92 & 12.56 & 2.90 & 12.72 & 2.90 & 12.83 & 2.94 & 13.18 & 2.77 \\ 
RoBERTa-B-L12    & \textbf{13.23} & \textbf{2.97} & \textbf{12.84} & \textbf{2.94} & \textbf{12.88} & \textbf{2.94} & \textbf{12.99} & \textbf{2.94} & \textbf{13.42} & \textbf{2.82} \\ 
\hline
\end{tabular}
}

\vspace{0.5em} 
\footnotesize 
* MAR and ** CTC don't rely on external pretrained models
\label{monolingual}
\end{table*}

%
\textbf{When using  PSLM as external knowledge source},

\begin{itemize}
    \item The improvements achieved by discrete tokens and continuous features are comparable, as observed from the two versions of HuBERT-L-L24.  
    \item Within the PSLMs, WavLM Base and HuBERT Large exhibit similar performance. However, scaling WavLM Base to WavLM Large leads to a significant further improvement.  
\end{itemize}

\textbf{When using PLM as external knowledge source},
\begin{itemize}
    \item RoBERTa-B-L12 outperforms all PSLMs, demonstrating the effectiveness of applying linguistic constraints supervision directly from the textual modality.  
    \item To further validate the contribution of PLM, an additional baseline is introduced, where the original outputs are jointly optimized using CTC loss. Since CTC optimization can also be interpreted as introducing direct textual supervision, its impact is analyzed.  
    \item The results show that CTC optimization significantly enhances performance, even surpassing all PSLMs. However, a performance gap remains compared to PLM, further validating the effectiveness of our method.  
    \item Optimization with linguistic knowledge from PLM is more effective at a low cost. MAR and AVSepChain enhance performance through improved linguistic modeling abilities, but both incur additional inference costs.  MAR requires three additional transformer encoder layers and two linear projection layers~\cite{wu2024target_cvpr}, while AVSepChain needs one cross-attention layer and three Conv1D layers, and integrates AV-HuBERT~\cite{AVhubert} as the visual frontend~\cite{avsepchain} to extract visual cues, also integrates HuBERT~\cite{HuBERT} as backend to resynthesis predicted speech. In contrast, optimization with linguistic knowledge required no extra cost during inference and could perform best with RoBERTa-B-L12. 

\end{itemize}

\vspace{-5pt} 
\subsection{Evaluation on the Monolingual Test Set}


The results are shown in Table. \ref{monolingual}. We observe the following: AV-Sepformer demonstrates relatively modest performance on all five monolingual test sets compared to its performance on the core test set, despite the model being exposed to all five languages. In particular, the performance for PT, the SI-SDR is 11.35, and for ES, the SI-SDR is 11.38, likely due to these languages having a small proportion in the training set. Notably, while English accounts for the majority of the training set, its monolingual performance is lower than that of the core set. Mixtures of the same language may pose more challenges for extractions.

With linguistic knowledge from RoBERTa-B-L12,  model consistently performs best across all five monolingual test sets. For PT and ES, improvements of approximately 1.5 dB in SI-SDR were achieved without additional fine-tuning. It is promising and can be extended to even lower-resource languages, where collecting multimodal data poses significant challenges.

\subsection{Evaluation in Visual Cue-Impaired Scenarios}
\vspace{-5pt}
In this section, we investigate whether linguistic constraints also help when primary visual cues are impaired. We directly evaluate the AV-TSE models with visual cue-impaired test set, including three visual cue-impaired scenarios: full visual missing, partial visual occlusion, and low resolution. For a fair comparison, we also employ ImagineNET as a comparison baseline, an AV-TSE system especially tailored for the visual cue missing scenario, by explicitly recovering target visual cue embedding with several visual decoders \cite{ImagineNET}. 
We employ two model settings: ImagineNET and ImagineNET-Sys.2, the former is re-trained on our AV-TSE training set, while the latter is from origin checkpoints, which has been trained on the visual full missing dataset.



 \begin{figure}[!htb]
\centering
\includegraphics[scale=0.35]{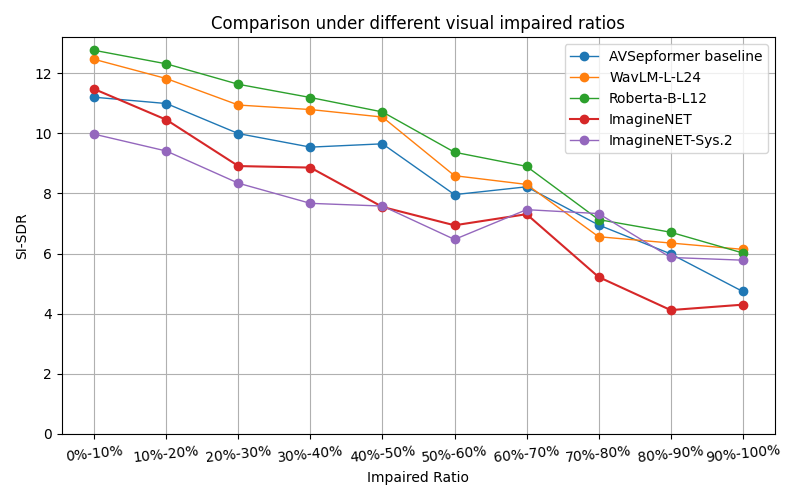}
\vspace{-5pt}
\caption{SI-SDR comparison under different visual cue-impaired ratios.}
\label{v_occ}
\end{figure}

\vspace{-5pt}
Specifically, we compute the average performance on three visual impaired scenarios with impairment ratios ranging from $0\%$ to $100\%$, as illustrated in Figure. \ref{v_occ}. We observe that RoBERTa-B-L12, and WavLM-L-L24 consistently outperform the vanilla AV-Sepformer across nearly all impairment ratios, and RoBERTa-B-L12 still achieves the best performance. In contrast, ImagineNET demonstrates modest performance at lower impairment ratios, but with high impaired ratios, it shows comparable performance to RoBERTa-B-L12 and WavLM-L-L24. This is likely due to its training bias towards fully missing visual scenarios, as well as model size limitations.

\begin{table}[ht]
\centering
\vspace{-5pt}
\caption{SI-SDR performance under three visual impaired scenarios, ``Full occ'' denotes full visual missing, ``Part occ'' denotes partial visual occlusion, ``Low reso'' denotes low resolution, and Avg denotes average performance on three scenarios.}
\vspace{-5pt}
\begin{tabular}
{p{0.16\textwidth}p{0.04\textwidth}p{0.04\textwidth}p{0.04\textwidth}p{0.04\textwidth}} 
\hline
\textbf{Knowledge Source} & \textbf{Full occ} & \textbf{Part occ} & \textbf{Low reso} & \textbf{Avg} \\ 
\hline
- (baseline)        & 5.15 & 11.26 & 10.72 & 9.04 \\ 
ImaginNET           & 2.92 & 11.38 & 10.40 & 8.22 \\ 
ImaginNET-Sys.2 \cite{ImagineNET}           &  5.36 & 9.75 & 10.40 &  8.01 \\ 
\hline
WavLM-L-L24         & 5.51 & 12.47 & 12.00 & 9.98 \\ 
RoBERTa-B-L12 & \textbf{6.07} & \textbf{12.87} & \textbf{12.32} & \textbf{10.42} \\ 
\hline
\end{tabular}
\label{vocc_table}
\vspace{-10pt}
\end{table}
The overall SI-SDR performance of different AV-TSE models on three visual cue-impaired scenarios is reported in Table. \ref{vocc_table}. Note that AV-Sepformer with RoBERTa-B-L12 or WavLM-L-L24 were not exposed to any visual cue-impaired scenario, but still outperformed ImagineNET Sys.2 across all impair ratios, where the latter have seen visual full missing conditions during training. This finding suggests the proposed linguistic constraints' practical applicability in real-world scenarios.

\vspace{-5pt}

\subsection{Cross-dataset Evaluation on LRS3}
In this section, we directly evaluate the proposed models on the LRS3 test set for cross-domain evaluation. The LRS3 dataset is collected from TED videos, containing less background noise and music compared to VoxCeleb2~\cite{Afouras2018LRS3TEDAL}, and it is widely used in visual speech recognition tasks \cite{Hong2023IntuitiveMA}. We select knowledge sources that perform best on speech and text modality, respectively, i.e. WavLM-L-L24 and RoBERTa-B-L12 for evaluation.
We also compare our results with AVSepChain on the LRS3 test set, as reported in the original paper using two-second speech segments \cite{avsepchain}.
The overall results are shown in Table. \ref{LRS3}.
\begin{table}[ht]
\centering
\vspace{-5pt}
\caption{Cross domain evaluation on LRS3 test set.}
\vspace{-5pt}

\begin{tabular}
{p{0.16\textwidth}p{0.08\textwidth}p{0.06\textwidth}} 
\hline
\textbf{Knowledge Source} & \textbf{SI-SDRi } & \textbf{PESQ } \\ 
\hline
- (baseline)   & 13.83 & 2.77    \\ 
AVSepChain \cite{avsepchain}   & 15.3 & \textbf{3.26} \\
\hline
WavLM-L-L24   & 15.21 & 2.99 \\ 
  
RoBERTa-B-L12 & \textbf{15.42}   & 3.04 \\ 
\hline
\end{tabular}
\label{LRS3}
\end{table}

We observe that all models achieve better performance compared to the results on Voxceleb2, likely due to fewer noise interferences. Furthermore, RoBERTa-B-L12 achieves the best performance in terms of SI-SDRi, demonstrating the effectiveness of linguistic constraints

\vspace{-10pt}
\section{Conclusion}
\vspace{-3pt}
In this study, we propose to derive linguistic constraints from pre-trained language models and speech-language models to optimize AV-TSE. Evaluated on multilingual, visual cue-impaired, and cross-domain test sets, the proposed paradigm consistently outperforms the vanilla AV-TSE backbone, while the linguistic constraints from the language model surpass those of the speech model.  
In future work, we aim to explore other methods of injecting language information and semantics to further improve AV-TSE performance.

\section{Acknowledgements}
This work was supported by National Natural Science Foundation of China, Grant No. 62401377, Shenzhen Science and Technology  Program (Shenzhen Key Laboratory, Grant No. ZDSYS20230626091302006), Shenzhen Science and Technology Research Fund (Fundamental Research Key Project, Grant No. JCYJ20220818103001002),
Program for Guangdong Introducing Innovative and Entrepreneurial Teams, Grant No. 2023ZT10X044, Yangtze River Delta Science and Technology Innovation Community Joint Research Project (2024CSJGG01100).

\bibliographystyle{IEEEtran}
\bibliography{mybib}

\end{document}